\documentstyle[twoside,fleqn,espcrc2,epsfig]{article}

\newcommand{\Ns}{N_{\sigma}}
\newcommand{\Nt}{N_{\tau}}

\newcommand{\AmS}{{\protect\the\textfont2
  A\kern-.1667em\lower.5ex\hbox{M}\kern-.125emS}}

\title{The string tension in SU(N) gauge theory from a careful analysis of 
smearing parameters.}

\author{C. Legeland with B. Beinlich, M. L\"utgemeier, A. Peikert and
  T. Scheideler \\ \vspace {1ex} 
Fakult\"at f\"ur Physik, Universit\"at Bielefeld, 
    Postfach 100131, D-33501 Bielefeld}
  
\begin{document}

\begin{abstract}
We report a method to select optimal smearing parameters before production runs
and discuss the advantages of this selection for the determination of the
string tension.
\end{abstract}

\maketitle

\section{Motivation and Introduction}

Even though computer power rises it is important to perform the measurements
and the data analysis in an optimal way.
In the case of the string tension this means that we want to gain maximum
information from Wilson loop measurements in minimal time.
The smearing procedure proposed by Albanese {\it et al.}\cite{Albanese87}
is an important contribution in this direction.

In principle everything is clear. Select smearing parameters, produce data
and analyse/extract the string tension. But what smearing parameters to choose?
Measuring a wide range of smearing parameters is expensive; but does it
improve the result?

We wanted one set of smearing parameters per coupling to keep CPU costs as
low as possible.

Focused on the long distance string behaviour of the potentials measured,we
face two problems in the analysis: which local potential to take and which fit
ansatz to the potential gives us the string tension? 

We show that in each of the steps towards the string tension 
choices and systematic errors made in the analysis can significantly change
the value of the string tension without increasing the statistical error 
correspondingly. This results in small statistical errors but large
systematical errors, that are difficult even to estimate.

We propose a standard procedure, based on physical arguments that minimizes
these systematical errors. We have tested this method for different $\beta$,
dimension, group, improvement e.g. 1x2, 1x2 with tadpole and also with
fermions, but for simplicity here all our results are shown for $\beta =14$
in 3d pure gauge SU(3).

One general remark: 
Smearing as proposed in \cite{Albanese87}
improves the signal to noise ratio significantly and does not affect the 
value of $\sigma$, 
iff the same parameters are applied to the Wilson loops at given $R$ and to all
configurations.

\section{Smearing}

The smearing procedure replaces a spatial link with a sum of the link 
and $\epsilon$ times it's spatial staples. 
%\begin{equation}
%  U = U + \epsilon \sum {\rm staples (spatial) }.
%\end{equation}
This is applied to all links on the lattice and repeated $n$ times.
   
As test operator to fix $n$ and $\epsilon$, we selected the 
loop $W(R=N_{\tau}/2, T=1) = {\rm T} (n, \epsilon)$, because 
we found this Wilson loop maximal improved (lifted ``from noise'').

For a given coupling $\beta$ we examine T$(n, \epsilon)$ 
for $n = 1, ... 40$ and $\epsilon = 0.02, ... 1.0$.\\
\centerline{\epsfig{file=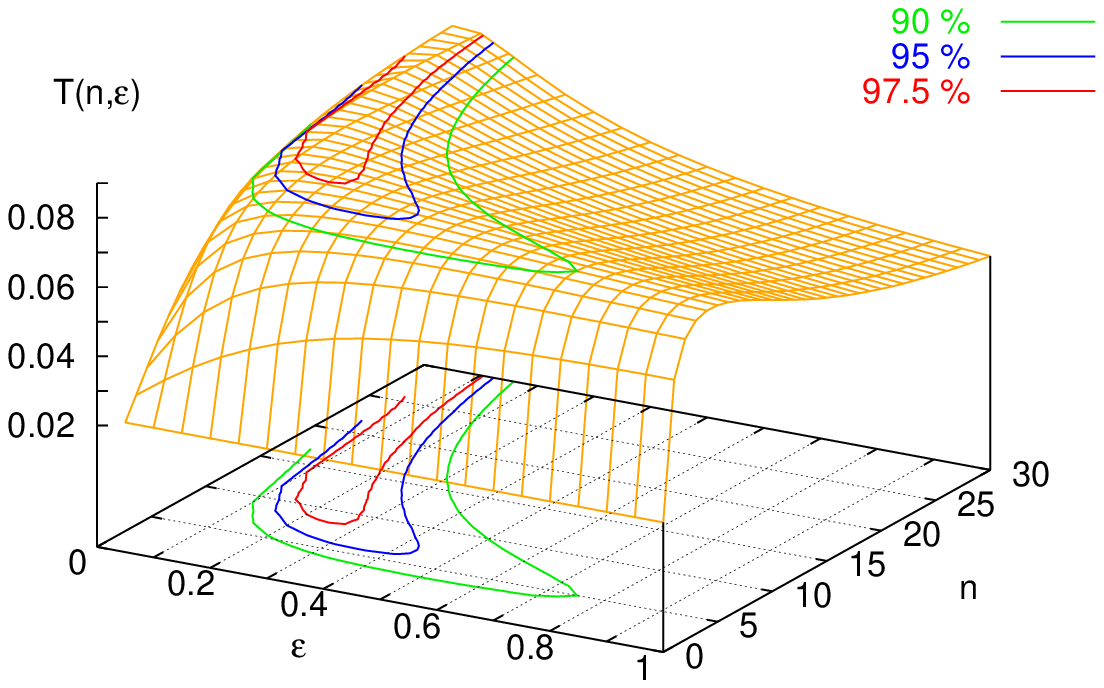, width=70mm } }
\stepcounter{figure}
Figure 1. Test operator T$(n, \epsilon)$ vs. n and $\epsilon$ with the
isolines of 90\%, 95\% and 97.5\% of maximal improvement.

\vspace{1ex}

From figure 1 we see that we reach 97.5\% of the
maximal possible signal with a set of smearing parameters:
$n=10, \epsilon = 0.2$ for this $\beta$. 
One should keep in mind that this figure is different
for different values of $\beta$, dimension, group, improvement
and fermions. Scanning over a larger area of $n$ and $\epsilon$ we find the
isolines closed. Smearing to much can make the signal worse than before.

\section{Potential}

The local potential $V_T(R)$
\begin{equation}
  V_T(R) = \log \frac{\langle {\rm W}(R,T) \rangle}{\langle {\rm W}(R,T+1) \rangle},
\end{equation} 
gives the potential $V(R)$ in the limit of large $T$
\begin{equation}
  V(R) = \lim\limits_{T\rightarrow\infty} \; V_T(R).
\end{equation}

Because the local potential approaches the asymptotic value exponentally,
we will find significantly different potentials, depending on whether one
assumes that asymptotic behaviour sets in at $T=2, 3$ or 4.

\begin{figure}[thbp]
  \vspace{-6mm}
  \epsfig{file = 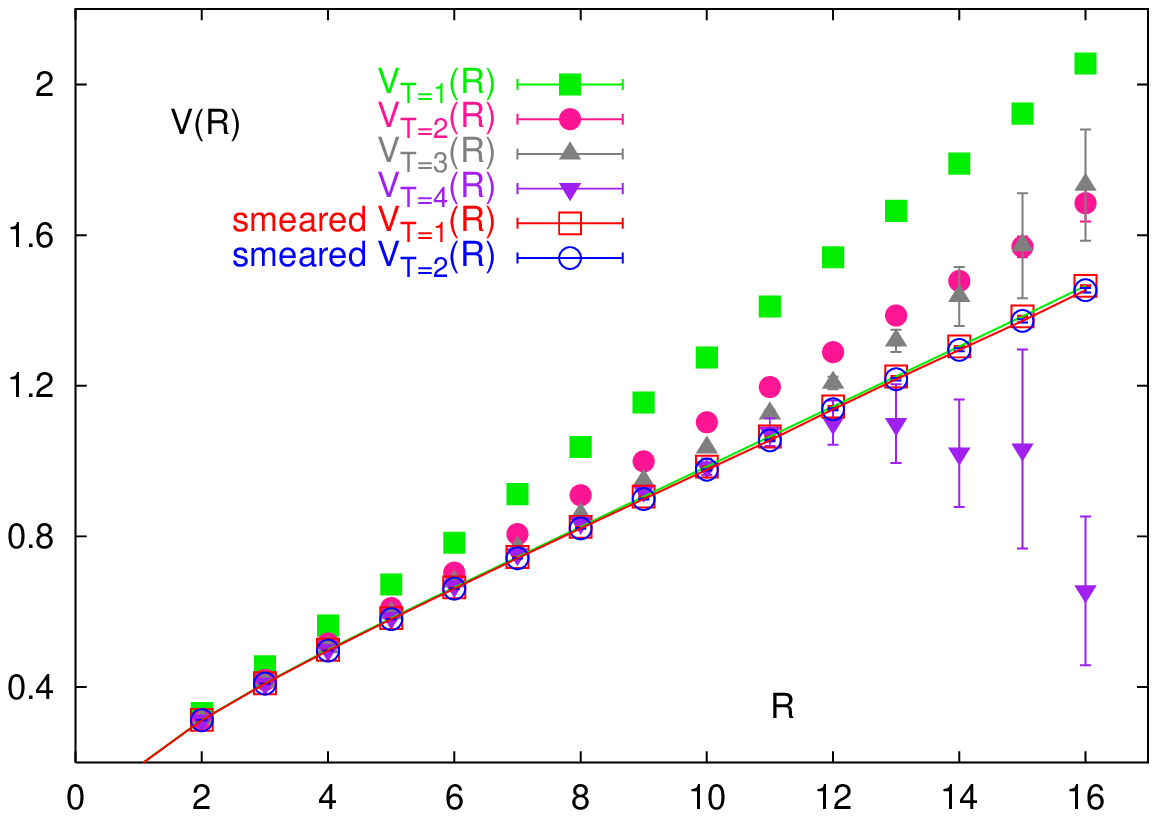, width = 74mm} 
  \vspace{-5mm}
  \caption{ Potentials $V(R)$ from different local potentials.} 
  \label{fig:VR}
  \vspace{-5mm}
\end{figure}

In figure \ref{fig:VR} we plot smeared and unsmeared potentials 
for $\beta =14$ in 3d SU(3).
The notation $V_{T = 2}$ means that the value for the potential $V(R)$ is
taken from the $V_{T = 2}(R)$. The unsmeared potentials decrease with increasing
T. Even $V_{T = 3}(R)$ is remarkingly higher than both smeared
potentials. The next potential taken from $V_{T = 4}(R)$ vanishes at long
distance in noise. (In \cite{martin} we show that similar effects are observed,
if the potential is smeared, but sub optimal smeared.) 
The smeared potentials are the same within errors, but there too, $V_{T = 2}(R)$
is lower (For $T\ge3$ smeared potentials are the same but errors grow). For the
fit we take the lowest, smeared $V_T(R)$, typically T = 2 or 3.

\section{Fit}
 
In (2+1) dimensions the coulombic force is logarithmic.
\begin{equation}
  V_{fit} = V_0 - \alpha \log R + \sigma R.
\end{equation}
Fits to this ansatz are unstable especially for large R. This suggests that
because of confinement
there is no long range coulombic content in the potential.

We then instead make  the Ansatz:
\begin{equation}
  V_{fit} = V_0 - \alpha / R + \sigma R.
\label{form:fit}
\end{equation}
This is motivated by the existence of a 
string fluctuation term $\frac{\pi (D-2)}{24 R}$ \cite{luescher}.
We fix $\alpha$ to the corresponding value 
and make fits from moderate R$(=R_{min})$ to $R=\Ns/2$.
Letting $\alpha$ free and find $\alpha \ge \frac{\pi(D-2)}{24}$
and the string tension from these fits larger than the above.

\begin{figure}[htbp]
  \vspace{-6mm}
  \epsfig{file = 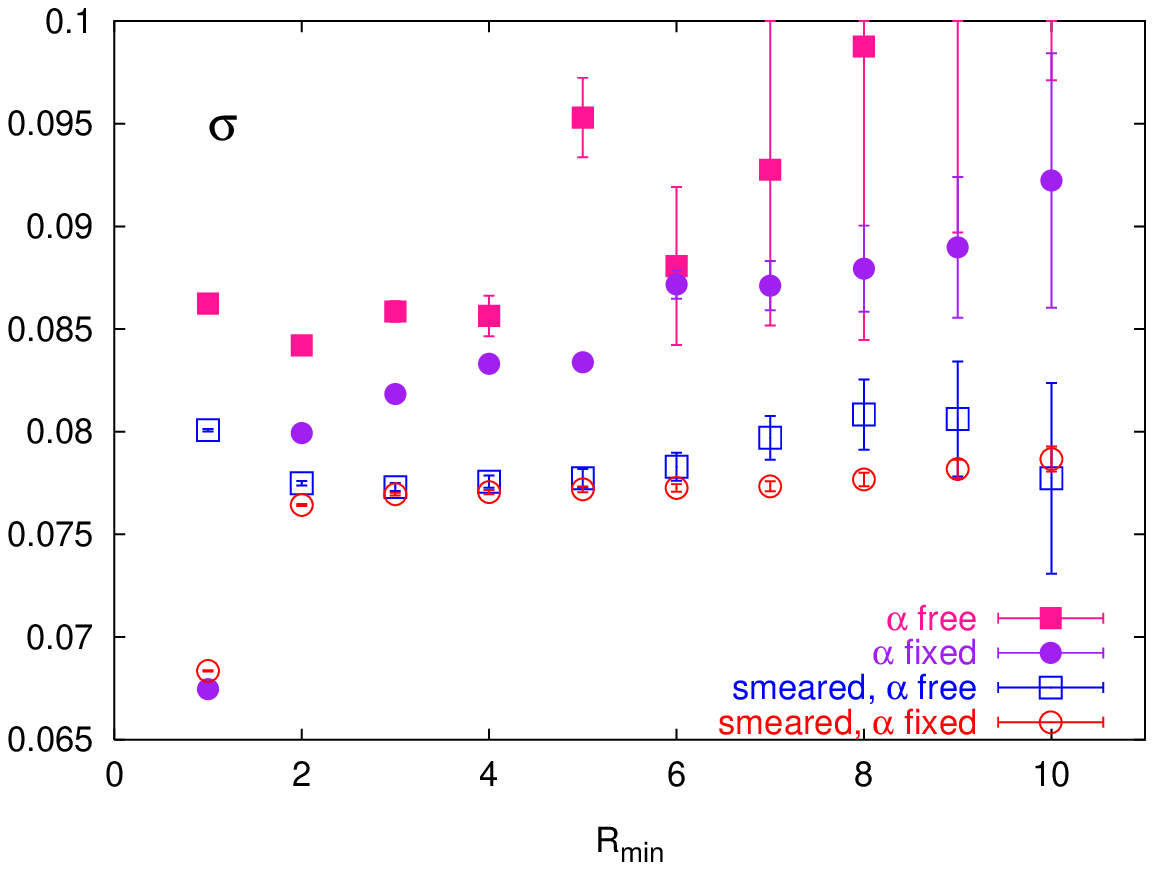, width = 74mm} 
  \vspace{-5mm}
  \caption{$\sigma$ vs $R_{min}$ from the fits to formula \ref{form:fit}}
  \label{fig:sigma}
  \vspace{-5mm}
\end{figure}

In figure \ref{fig:sigma} we plot the string tensions for different fit ranges
for free and fixed $\alpha$ fits. We see clearly that unsmeared potentials
result in unstable and increasing $\sigma$ for long distance fits.
The $\sigma$ for smeared potentials are stable and the same within errorbars
for all fits $R_{min}>3$, if $\alpha$ is fixed and only one standard deviation
from each other, if $\alpha$ is free.  
As we want to extract the long distance behaviour, we take 
$\sigma$ from the fit with $R_{min}= 4$ or $5, 6, 7$.

%\subparagraph{

\vspace{1ex}

{\bf Scaling of the string tension}

%\vspace{1ex}

In 3d we then plot $\beta \sigma$ versus $1/\beta$ and make a linear or
quadratic fit to these points. With these fits we calculate
e.g. $T_c/\sqrt{\sigma}$ at $\beta_{c}(\Nt, \Ns =\infty)$ \cite{martin}.

\section{Results: $T_c/\sqrt{\sigma}$ in 3d and 4d }

\subsection{\bf 3d}
\begin{figure}[htbp]
  \vspace{-6mm}
  \centerline{\epsfig{file = 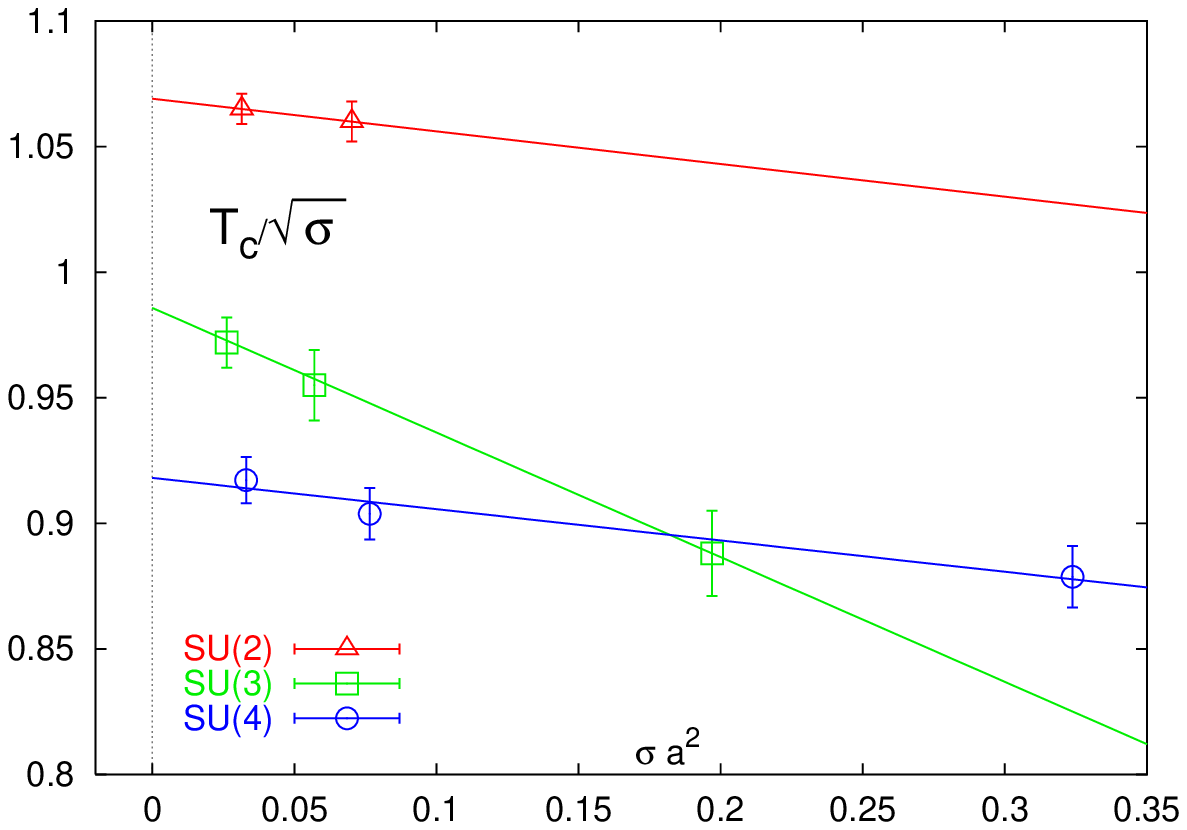, width = 74mm}}
  \vspace{-5mm}
  \caption{$T_c/\sqrt{\sigma}$ in 3d for SU(N), N=2, 3 and 4. The
    extrapolations to continuum are linear fits in $\sigma a^2$.}
  \label{fig:tcs3d}
  \vspace{-5mm}
\end{figure}
Our investigation is precise enough to clearly see that the SU(2), SU(3) and
SU(4) continuum values differ significantly. For SU(2) earlier results 
from M. Teper are of the same order but larger \cite{teper}. SU(3) meets within
errors the prediction (0.977) of the Nambu-Goto string
model\cite{olesen85}. SU(2) and SU(4) are only qualitativly similar to the
string model prediction.

\subsection{\bf 4d}
\begin{figure}[htbp]
%  \vspace{-5mm}
  \centerline{\epsfig{file = 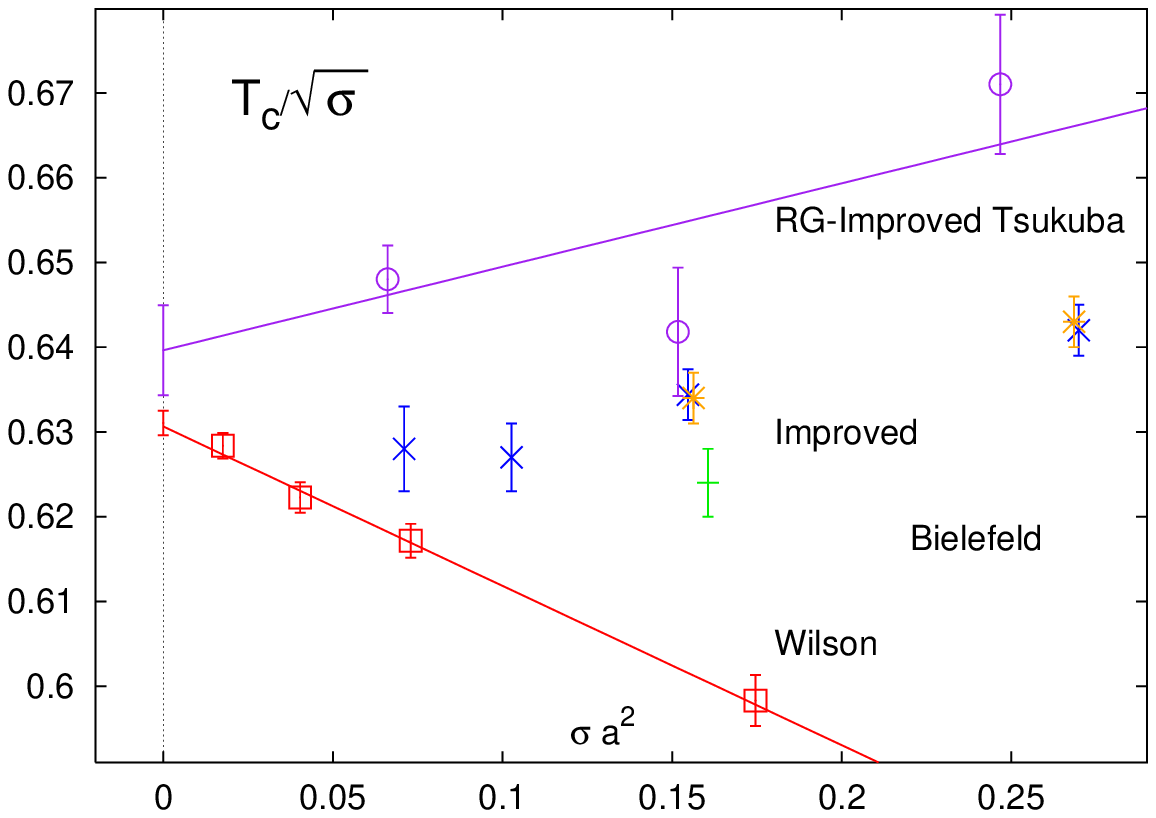, width=74mm}}
  \vspace{-5mm}
  \caption{$T_c/\sqrt{\sigma}$ in 4d for SU(3).The extrapolations to continuum
    are linear fits in $\sigma a^2$.}
  \label{fig:tcs4d}
  \vspace{-5mm}
\end{figure}

In figure \ref{fig:tcs4d} we show our pure Wilson, 1x2, 1x2 tadpole and
2x2 improved action results. 
In addition we plot our fits to the potentials 
published by Iwasaki{\it et al.}\cite{iwasaki}, labeled ``RG Improved
Tsukuba''. Contrary to Iwasaki {\it et al.} we fit the potentials with fixed
$\alpha$ and large $R_{min}$, this results in a larger $\sigma$, therefore 
lower $T_c/\sqrt{\sigma}$. 

A problem in (3+1) dimensions is, that both string fluctuation term
and coulomb forces have $1/R$ shape. A coulombic behaviour can not be ruled out
for long distances as easily as in 2+1 dimensions. We find in fits to formula
\ref{form:fit} that $\alpha$ is more consistent with a string fluctuation term,
because $\sigma$ becames more stable and less dependent from the fit range
choosen. 

\vspace{1ex}

{\bf Conclusions:}

If we apply the same analysis scheme to all actions, 
the continuum extrapolations of $T_c/\sqrt{\sigma}$ from our data 
are the same within errors.

In this analysis scheme the continuum string tensions from our fits to the
potential data from Iwasaki {\it et al.} and our calculations then differ
only by one standard deviation. 

\vspace{1ex}

The work has been supported by the DFG under grants Pe 340/3-3 and Pe 340/6-1.

\end{document}